\documentclass{aa}

\usepackage{graphicx}
\usepackage{txfonts}
\usepackage[colorlinks=true, citecolor=blue, linkcolor=black, urlcolor=black]{hyperref}

\usepackage{natbib}
\usepackage{amssymb}
\usepackage{color}
\usepackage{dcolumn}

\newcolumntype{d}{D{.}{.}{-1}}

\begin{document} 

  \title{Spins and shapes of asteroids reconstructed from photometry from surveys}
  
  \titlerunning{Spins and shapes of asteroids reconstructed from photometry from surveys}

  \author{J. \v{D}urech		\inst{1}	\and
	       J. Hanu\v{s}        \inst{1}	
	 }

  \institute{Charles University, Faculty of Mathematics and Physics, Institute of Astronomy, V Hole\v{s}ovi\v{c}k\'ach 2, 180\,00 Prague, Czech Republic\\
             \email{durech@sirrah.troja.mff.cuni.cz}
	     }

  \date{Received ?; accepted ?}

  \abstract
{Modern asteroid surveys routinely acquire sparse photometric measurements for hundreds of thousands of asteroids. These data can be exploited to reconstruct asteroid spin states and shapes, providing statistical constraints on the rotational evolution of different dynamical populations.}
{Our goal is to derive spin-state and shape models from publicly available survey photometry and use the enlarged sample to investigate the distributions of asteroid physical properties across the Solar System.}
{We combined sparse photometry from Catalina, Mt. Lemmon, Pan-STARRS, ZTF, USNO, ATLAS, ASAS-SN, TESS, and Gaia and applied convex light-curve inversion to determine sidereal rotation periods, spin-axis orientations, convex shapes, phase-function parameters, and colors. The extensive parameter-space search was performed using the distributed-computing project Asteroids@home.}
{We derived approximately 30\,000 asteroid models, including about 22\,000 unique spin-state solutions and almost 10\,000 partial pole solutions, tripling the number of available asteroid models. The enlarged sample confirms the dependence of the rotation-period distribution on taxonomic class, reveals a pronounced lack of elongated asteroids near the spin barrier, and shows that slow rotators exhibit systematically larger photometric residuals, supporting the interpretation that many are in non-principal-axis rotation. We derived spin-state distributions for 47 near-Earth asteroids, 247 Jupiter Trojans, and 157 Hildas. The expanded Trojan sample no longer shows a statistically significant excess of prograde rotators, indicating that the asymmetry reported from the smaller sample is weaker than previously suggested.}
{The combination of modern survey photometry and large-scale light-curve inversion enables population studies based on tens of thousands of asteroid models. The resulting database provides improved constraints on the rotational evolution of asteroids and constitutes a valuable resource for future studies of asteroid physical properties.}


  \maketitle

  \nolinenumbers

  \section{Introduction}  

    The number of known asteroids is rapidly growing as sky surveys continue to detect and observe new objects. The current count is about 1.5 million known asteroids, and an order-of-magnitude increase is expected soon with the launch of the Legacy Survey of Space and Time (LSST) at the Vera Rubin Observatory \citep{Kur:25}. Surveys provide not only astrometric measurements of asteroids needed to compute orbits but also photometric measurements, which are reported together with astrometry to the Minor Planet Center (MPC). This photometry contains invaluable information about the asteroids' shapes and spin states, as these properties affect the objects' brightness as a function of geometry and rotation phase. One of the methods for reconstructing an asteroid's spin and shape from photometry is the light curve inversion method \citep{Kaa.Tor:01, Kaa.ea:01}, specifically used in sparse-in-time photometry \citep{Kaa:04}. Over the last decade, the method applied to sparse photometry has led to models of approximately ten thousand asteroids \citep{Dur.ea:16, Dur.ea:18c, Dur.Han:18, Dur.ea:19, Han.ea:21, Dur.Han:23, Cel.ea:24, MacL.ea:26}. The sample of available asteroid models is small compared to the total number of known asteroids; yet, it provides information on how spin properties are distributed across the population. 

    This paper aims to increase the number of asteroid models further. We report the application of the light curve inversion method to combined photometric data, yielding approximately 30,000 new models. 

  \section{Photometric data}    

    As the source of asteroid photometry, we used mainly the photometry reported by observatories to the MPC.\footnote{\url{https://www.minorplanetcenter.net/}} The data are also available via AstDyS\footnote{\url{https://newton.spacedys.com/astdys2}}, from which we retrieved them. Namely, we downloaded {\tt rwo} files in February 2024 for the first 660,000 numbered asteroids. We then extracted photometric data observed before 2024 by these observatories: Catalina, Mt.\,Lemmon, Pan-STARRS 1 and 2, Zwicky Transient Facility (ZTF), and U.S. Naval Observatory (USNO). We used only photometry reported to two decimal places. 
    
    Additionally, we also used photometry directly downloaded from other repositories. These included: Gaia data release 3 \citep{Tan.ea:23}, available through the Gaia archive\footnote{\url{https://gea.esac.esa.int/archive/}} (limited to asteroids with at least ten photometric points), as well as photometry from the ATLAS Solar System Catalog, version 2 \citep{Ton.ea:18}\footnote{\url{https://astroportal.ifa.hawaii.edu/atlas/sscat/}},  All-Sky Automated Survey for Supernovae \citep[ASAS-SN,][]{Han.ea:21}, and TESS \citep{Pal.ea:20}, provided that each source contributed at least 20 points. The surveys used are listed in Table~\ref{tab:observatories}.

     \begin{table*}
        \caption{List of observatories and asteroid observations.}
        \label{tab:observatories}
        \centering
        \begin{tabular}{l c c r r c}
        \hline\hline
        Observatory     & MPC code              & filter    & asteroids & measurements    & years         \\
        \hline
        Catalina        & 703                   & G         & 131,436   &  51       & 2020--2023    \\  
        Mt.\,Lemmon     & G96                   & G         & 291,028   &  40       & 2006--2023    \\  
        Pan-STARRS 1, 2 & F51, F52              & w         & 281,598   &  46       & 2010--2023    \\  
        ZTF             & I41                   & g         &  76,375   &  37       & 2018--2022    \\  
                        &                       & r         & 112,786   &  48       & 2018--2023    \\  
        USNO            & 786                   & V         &     914   & 133       & 1993--2009    \\                          
        Gaia DR3        & 258                   & G         & 130,876   &  22       & 2014--2017    \\  
        ATLAS           & T05, T08, M22, W68    & c         & 294,151   & 133       & 2015--2024    \\
                        &                       & o         & 326,492   & 347       & 2015--2024    \\
        ASAS-SN         &                       & V         &  11,738   &  89       & 2012--2018    \\
        TESS            & C57                   & V         &   8,749   & 467       & 2018--2019    \\
        \hline        
        \end{tabular}
        \tablefoot{The table lists all observatories from which we obtained asteroid photometry for inversion. For each observatory and filter, the number of asteroids, the mean number of photometric measurements per asteroid, and the time span of observations are given.}
    \end{table*}

    After extracting photometry and epochs of observation, we computed the heliocentric and geocentric vectors for each photometric point using the Miriade ephemeris service at IMCCE\footnote{\url{https://vo.imcce.fr/webservices/miriade/}} and the JPL Horizons service.\footnote{\url{https://ssd.jpl.nasa.gov/horizons/}} For each asteroid, we combined all available data together, treating each observatory and each filter as a single internally calibrated light curve. We also removed outliers detected above the $3\sigma$ threshold during phase curve fitting \citep{Han.ea:11}.

    We did not account for the photometric errors of individual points, even when reported. Instead, we estimated mean errors of individual observatories the same way as \cite{Han.ea:23} by using available DAMIT\footnote{\url{https://damit.cuni.cz/}} shape models and data from individual observatories.
    We assumed that the root-mean-square residuals correspond to the photometric errors in the data and that these errors depend only on the mean apparent brightness. A quadratic function was then fitted to obtain an empirical formula for computing the error from the mean apparent brightness for each observatory and filter. The final dataset consists of 326,840 asteroids.

    Although some observatories use the same filters (e.g., Catalina, Mt.~Lemmon, and Gaia, which observe in the G filter; see Table~\ref{tab:observatories}) and such data could be combined into a single light curve, we treated them as independent light curves with an arbitrary shift between them. We assumed that internal calibration within a single observatory is more accurate than absolute calibration between observatories. By using this approach, we were not affected by possible systematic shifts between observatories \citep{Hof.ea:25}.

  \section{Modeling}  

    We followed the same approach as in our previous studies: \cite{Dur.ea:19, Han.ea:21, Dur.Han:23}. The aim was to find the global minimum in the parameter space and determine whether it was significantly lower than other local minima. To find the best model, we scanned the sidereal rotation period on the interval 2--10,000 hours. The lower limit corresponds to the observed spin-barrier of about 2.3\,h for asteroids larger than about 150\,m \citep{Pra.Har:00}; the upper limit is large enough to cover even extremely slow rotators with periods of thousands of hours detected recently by \cite{Era.ea:21}. For each trial period, the optimization was run with ten initial pole directions. The phase curve parameters were also optimized. 
    
    To find the globally optimal solution in the period parameter space, it is necessary to run the optimization for hundreds of thousands of trial periods, which requires significant computing time. To manage this task for more than 300,000 asteroids, we used the distributed computing project Asteroids@home \citep{Dur.ea:15}.\footnote{\url{https://asteroidsathome.net}} It uses the Berkeley Open Infrastructure for Network Computing (BOINC) framework to distribute computational units that are solved independently using individual computers belonging to volunteers connected to the project. With about 100 TeraFLOPS of computational power, the data were processed in about half a year.

    At the same time, we ran an alternative scan of the period parameter space using triaxial geometrically scattering ellipsoids. The advantage of this approach is that the disk-integrated brightness can be computed analytically \citep{Ost.Con:84}. This results in a speedup of approximately two orders of magnitude compared to the approach with general convex shapes. Ideally, the best-fit periods from convex shapes and ellipsoids are the same.

    To select only unique and reliable solutions, we followed the pipeline described by \cite{Han.ea:11} and \cite{Dur.ea:18c}. By combining periodograms from ellipsoids and convex shapes, checking for period consistency, and excluding solutions with exceedingly high residuals, we obtained models for 21,763 asteroids. We also verified whether the shape rotates around the principal axis. For another 9436 asteroids, we obtained partial models: the ecliptic longitude of their spin pole was not constrained, only the latitude.
    
    In approximately 70 cases, we find that excluding TESS data resulted in a unique model. This is due to the large errors in the TESS data, which spoil the solution. Due to the long 30\,min exposure time, TESS data should be modeled differently, as the brightness of a fast rotator can change significantly over this time. While this can affect the asteroids with the fastest rotation (2--3\,h) in our sample, longer rotation periods are practically unaffected \citep{Fen.ea:24}, and we neglected this effect in our approach.

    For rotation periods longer than 5000\,h, we accepted the solution only if it was supported by solutions from both the convex models and ellipsoids. This was to avoid false-positive solutions for extremely long periods. Also, the solution was accepted only if the time interval of observations lasted for at least 1000 days.

    The inversion algorithm also provides information about the phase curve. The linear-exponential function introduced by \cite{Kaa.ea:01} describes the phase-angle dependence of the light-scattering model. It has three parameters optimized during the inversion. The phase curve is assumed to be the same for all filters. The magnitude difference between filters defines the corresponding color.
    
    \begin{figure*}
        \sidecaption
        \includegraphics[width=12cm]{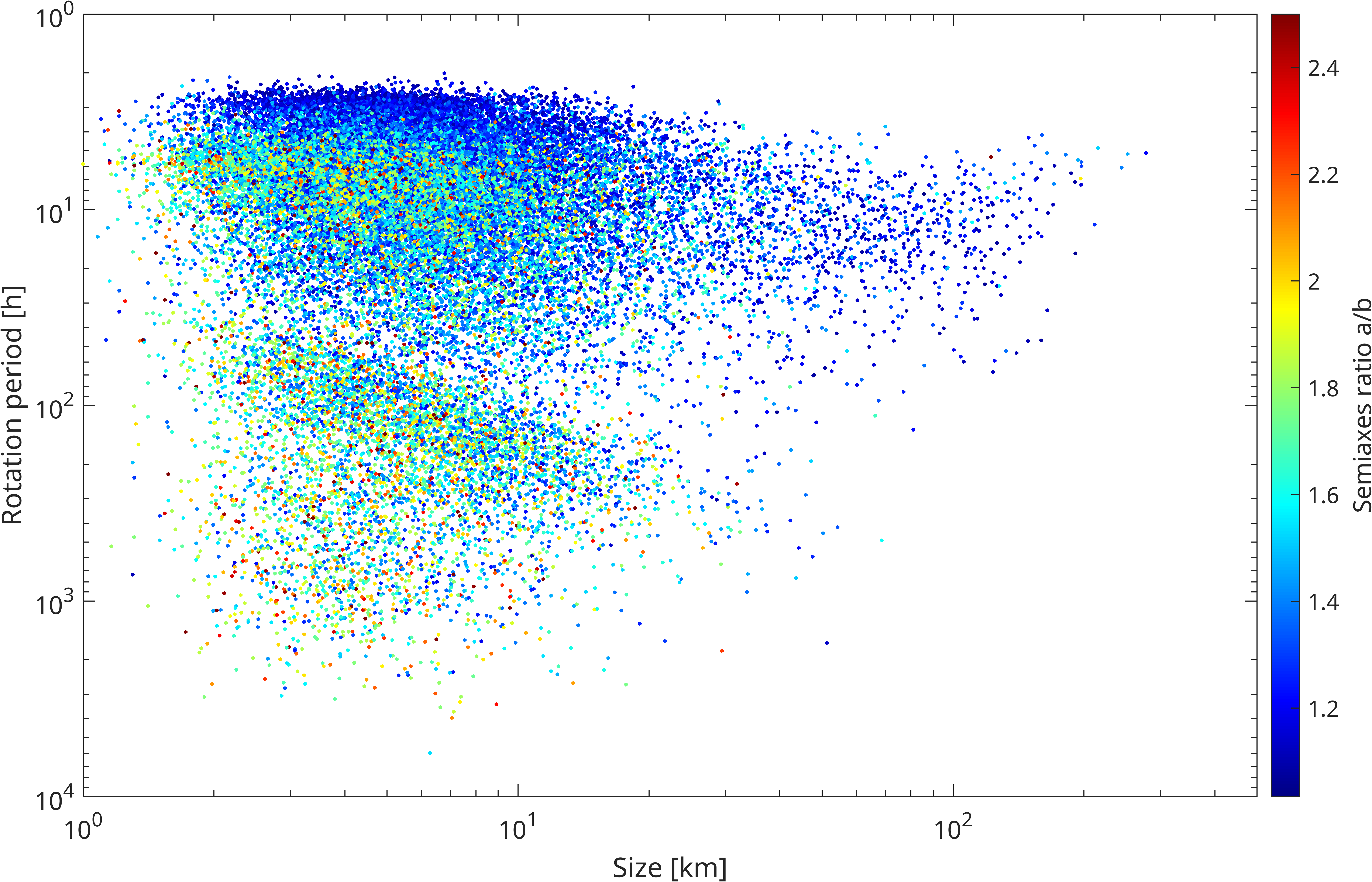}
        \caption{Period vs. size for bootstrap periods and the ellipsoidal model. The color represents the $a/b$ ratio of the best-fit ellipsoids. Near the spin barrier, only spheroidal shapes with small $a/b$ values are observed (blue). The $a/b$ values above 2.5 (the maximum was 4.0) were set to 2.5 to achieve better coverage of the color palette.}
        \label{fig:P_vs_D}
    \end{figure*}

    \subsection{Comparison with DAMIT models}

        We compared our results, namely the rotation periods, with the models from DAMIT, version from 3 October 2025. The overlap between our models and those in DAMIT was 6718 asteroids. We compared their rotation periods and find 163 cases in which the relative difference between the periods exceeded 1\,\% (Table~\ref{tab:DAMIT}). Most of the inconsistent DAMIT models were reconstructed from limited sparse photometry and come from the three sources: \cite{Dur.ea:18c},  \cite{Dur.ea:19}, and \cite{Dur.Han:23}. We update DAMIT records with the new models. The only exception is asteroid (10)~Hygiea, for which we derive a rotation period of 27.6589\,h, which is twice the correct period of 13.8256\,h \citep{Ver.ea:20}. Hygiea is almost spherical with significant albedo variegation, which produces only one maximum and one minimum per rotation on its light curves, contrary to the two maxima and minima for nonspherical asteroids.  
    
    \subsection{Rotation periods from the ellipsoidal model}  

        The total number of asteroids for which we derived either a full or partial model was about 30,000. To further expand the sample of asteroids with known rotation periods (without attempting to derive the full spin state), we followed the approach of \cite{Dur.ea:22}. This method assumes that if the best-fit period remains consistent even for randomly resampled data, it likely represents the true period of rotation. We generated bootstrap light curve data by randomly resampling each light curve block separately. We repeated this ten times for each asteroid. From the original periodograms with ellipsoids, we selected the best period, then selected those asteroids for which these ten periods were the same (within 1\%), as with the original dataset (68,259 asteroids). We required the standard deviation of the periods divided by the period to be lower than 0.05\% and the relative standard deviation of $a/b$ elongation to be lower than 20\,\% (55,299 asteroids). Then, we compared these periods with the best periods from the convex period search, using the 0.05\,\% limit for the difference between the periods (48,250 asteroids). These limits were set subjectively to filter out numerous false solutions at approximately 48\,h and 9000\,h. The results are shown in Fig.~\ref{fig:P_vs_D} for 32,238 asteroids with known sizes taken from the Minor Planet Physical Properties Catalogue (MP3C).\footnote{\url{https://mp3c.oca.eu/}} Because the period search was limited to the interval between 2 and 10,000 hours, no periods exist outside this interval. 

        We compared our bootstrap-derived periods with values reported in the Asteroid Lightcurve Database\footnote{\url{https://www.minorplanet.info/php/lcdb.php}} \citep[LCDB][]{War.ea:09}, using an uncertainty code $U = 3$ (i.e., the most reliable periods). The October 2023 LCDB release reports 4183 periods with $U = 3$. For 3150 of them, we obtain a corresponding bootstrap period. Of these, 3069 periods agree within 5\,\%, which is 97.4\,\%. According to a similar analysis by \cite{Dur.ea:22}, the number of false-positive periods is only a few percent. However, the LCDB sample is dominated by rotation periods shorter than 100\,h, with only about 60 longer periods. Hence, the frequency of false-positive solutions among slow rotators is not robustly constrained.

        In Fig.~\ref{fig:P_vs_D}, we can see that, near the spin barrier of 2 hours, asteroids are almost spheroidal, with low shape elongation $a/b$. Asteroids that are more elongated, with higher values of $a/b$, do not rotate as fast because their spin barrier is shifted toward longer periods \citep{Pra.Har:00, Was.ea:15}. 

        Another apparent feature in Fig.~\ref{fig:P_vs_D} is the valley with a lower concentration of asteroids, which separates slow rotators with rotation periods $\gtrsim 50$\,h from faster ones. The physical mechanism explaining this valley was proposed by \cite{Zho.ea:25} and predicts that most slow rotators are in the tumbling rotation state, to which they were excited by collisions. They remain there for a long time before dissipating energy, transitioning to the principal axis mode, and being accelerated to faster rotation by the YORP effect. Recently, \cite{Dye.ea:26} showed that the position of the valley is different for S and C taxonomic types, depending on the composition. For C-complex asteroids, the valley occurs at longer periods than for the S-complex asteroids. This is likely related to porosity and energy dissipation efficiency.
    
\begin{figure}
            \includegraphics[width=\columnwidth]{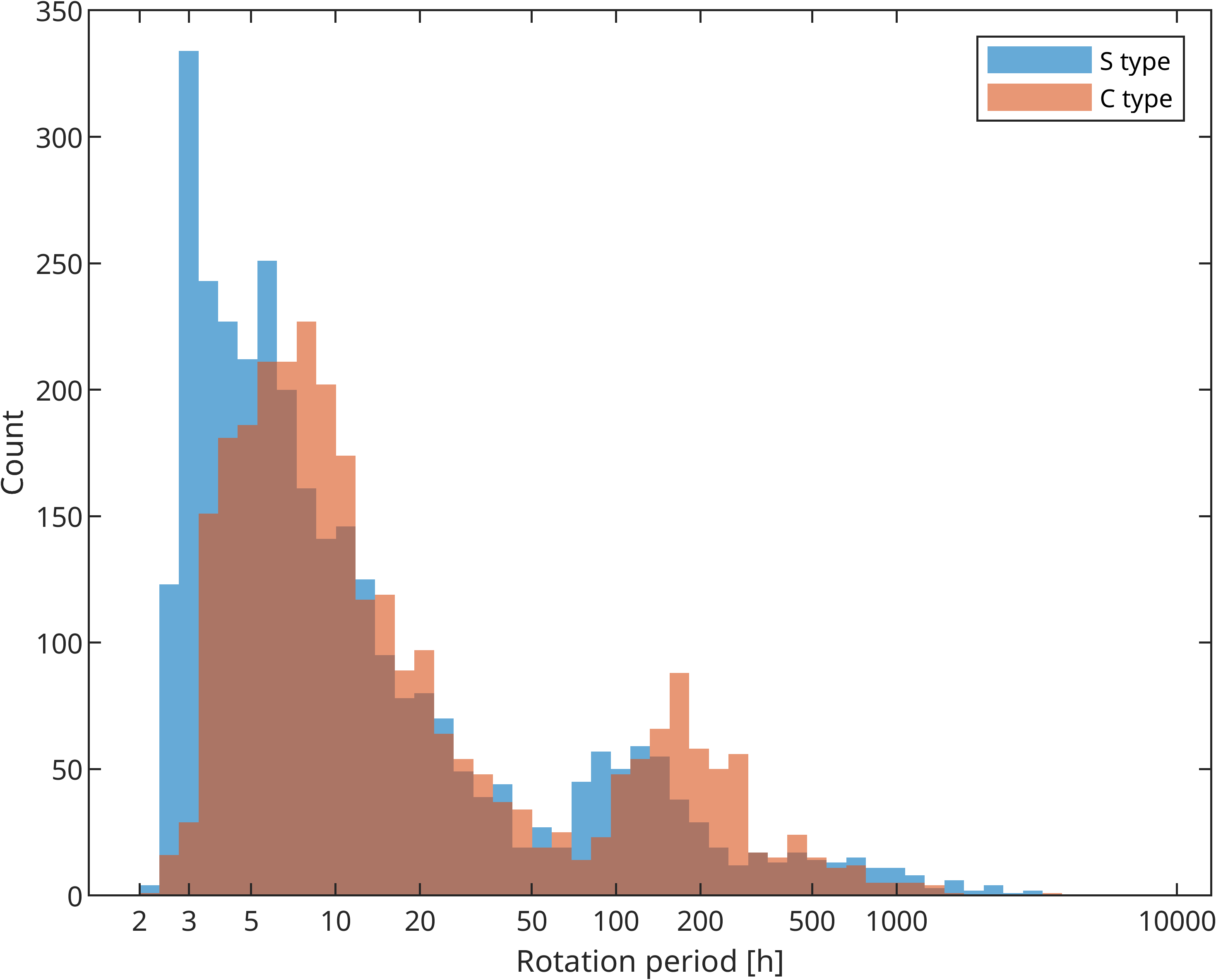}
            \caption{Period distribution for S- and C-type asteroids in the 5--15\,km size range. The peak of slow rotators is shifted toward longer periods for C-type asteroids. At shorter periods, S-type asteroids exhibit two peaks at 3 and 6\,h, while C-type asteroids exhibit only one peak around 8\,h.}
            \label{fig:P_histogram}
        \end{figure}

        We qualitatively see the same trends as \cite{Dye.ea:26}. The distribution of rotation periods for S- and C-type asteroids is significantly different, as shown in Fig.~\ref{fig:P_histogram} for the asteroids in the size range 5--15\,km. The density peak of slow rotators is shifted toward longer periods for C-type asteroids. The distributions are different also at shorter periods: while S-type asteroids have two peaks at 3\,h and 6\,h, C-type asteroids have one peak at about 8\,h. The spin barrier is shifted to about 3\,h for C-type asteroids, likely because of their lower densities. For asteroids smaller than about 5\,km, our model sample is strongly dominated by S-type asteroids, so we cannot compare their period distributions. The taxonomic classification was taken from the MP3C site. We included all asteroids with probabilities of belonging to the S or C class higher than 50\,\%.

    \subsection{Fast rotators}

    We limited our period search interval to 2--10,000 hours. The lower limit is justified by the fact that according to the LCDB, there are only about a dozen asteroids larger than 1\,km and rotating faster than the spin barrier of about 2.1\,h with a reliably determined period (U = 3- or 3). Limiting our period range prevents us from finding potential fast rotators. However, searching for them blindly without any preselection based on size would be time-consuming. This is because all local minima in the period parameter space must be tested and the distance between them decreases for shorter rotation periods \citep{Kaa:01}.

    For some known fast rotators in our sample, we expanded the tested period range to confirm their rotation periods. For example, asteroid (40511)~1999 RE88 has a rotation period of $1.96 \pm 0.01$\,h, according to \cite{Cha.ea:16}. Extending the period search interval to 1--10,000\,h, the best-fit period becomes 1.95217\,h. However, the minimum was not deep enough to pass our threshold. Another example is asteroid (44145)~1998 HJ101, for which \cite{Cha.ea:22} derive a rotation period of 1.74\,h. Our search over an interval starting from 1\,h yields the best unique period of 1.73758\,h. Several pole solutions are acceptable. The fastest rotator we tested is (267812)~2003 TX19, for which \cite{Cha.ea:22} report a rotation period of 0.48\,h. The best period we found on the interval 0.3--10,000\,h is 0.484150\,h.

    \subsection{Near-Earth asteroids}

\begin{figure}
        \includegraphics[width=\columnwidth]{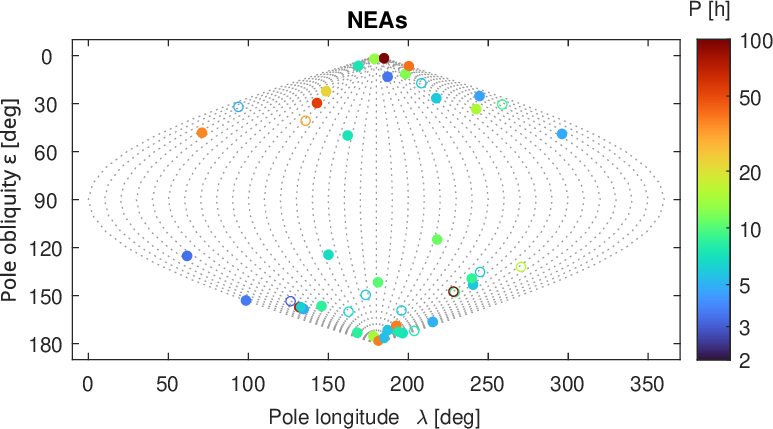}
        \caption{Spin and period distribution of NEAs. The filled circles represent full models (the first pole), and the empty circles are partial models (the best pole). The color of the circles corresponds to the rotation period.}
        \label{fig:NEAs}
    \end{figure}

    Our sample includes 881 Near-Earth asteroids (NEAs), for 47 of which we derived models. The distribution of their spins and periods is shown in Fig.~\ref{fig:NEAs}. We see clear clustering toward the extreme values of obliquity as well as asymmetric rotation. We identify 18 prograde and 29 retrograde rotators, consistent with the findings of \cite{LaS.ea:04}, who attributed the observed excess of retrograde rotators to a quantitative agreement with the theoretical predictions of the Yarkovsky model and delivery of NEAs from the main belt.
    
    \subsection{Trojans}

Our sample also contains photometric data for 2\,755 Jupiter
Trojans. We derived unique spin-state and shape models for 157
asteroids in the L4 swarm and 90 in the L5 swarm, increasing the
sample of unique Trojan models used for statistical analyses from 90
\citep{Han.ea:23} to 247. Although the overall completeness
corresponds to only $\sim$9\,\% of the currently known
Jupiter Trojans, it is substantially higher for the brightest objects
and decreases rapidly toward smaller sizes, reflecting the increasing
difficulty of deriving reliable spin-state solutions for faint
targets.

For objects common to the present sample and that of
\cite{Han.ea:23}, the derived rotation periods are mutually
consistent. The pole solutions generally retain the same rotation
sense, although individual pole longitudes can differ due to the substantially enlarged photometric datasets and the resulting
removal or modification of some mirror-pole solutions. This agreement
provides an internal consistency check and indicates that the change
in the inferred prograde-to-retrograde ratio is caused predominantly
by the addition of new objects rather than by systematic revisions of
the previously derived spin states.

\begin{figure}
        \includegraphics[width=\columnwidth]{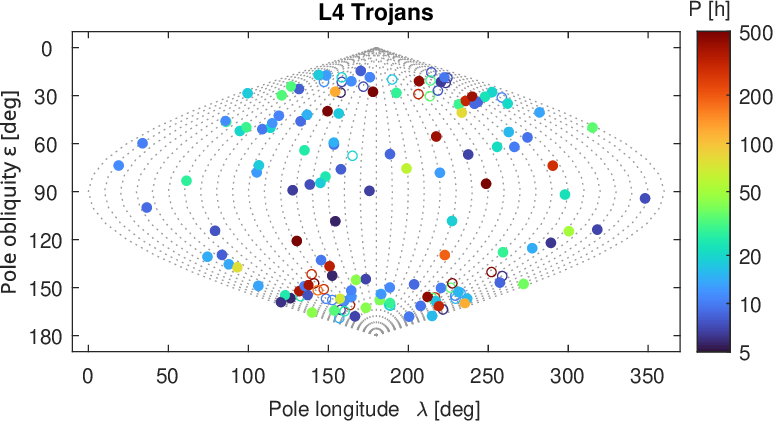}\\
        \includegraphics[width=\columnwidth]{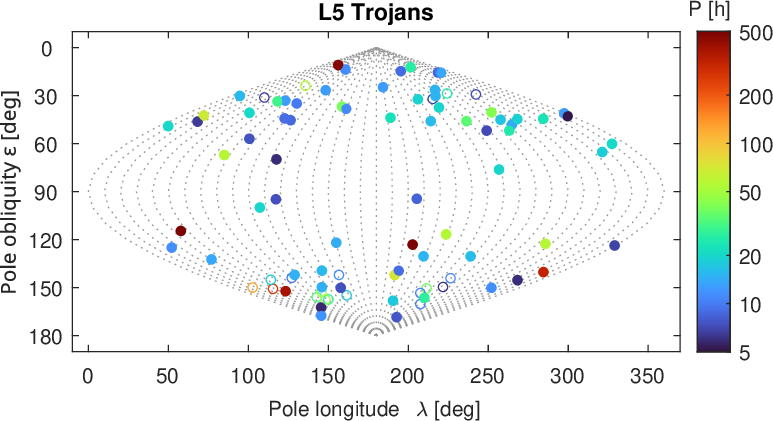}
        \caption{Spin and period distribution of Jupiter Trojans. The filled circles represent full models (the first pole), and the empty circles are partial models (the best pole). The color of the circles corresponds to the rotation period. Rotation periods longer than 500\,h are mapped to 500\,h.}
        \label{fig:Trojans}
    \end{figure}

The distribution of spin axes is shown in
Fig.~\ref{fig:Trojans}. As expected for objects at Jupiter's
heliocentric distance, the clustering of spin axes toward high
ecliptic latitudes is much weaker than for main-belt asteroids,
consistent with the reduced influence of the YORP effect. In contrast
to \cite{Han.ea:23}, we find no statistically significant excess of
either prograde or retrograde rotators. Likewise,
Kolmogorov--Smirnov and Mann--Whitney tests reveal no significant
differences between the L4 and L5 swarms in their distributions of
rotation periods and shape elongations, while Fisher's exact test
indicates that the fractions of prograde and retrograde rotators are
statistically indistinguishable between the two swarms. The shortest
rotation period in our sample is 5.2\,h.

The enlarged sample extends toward somewhat smaller Trojans than that
analyzed by \cite{Han.ea:23}. The newly modeled objects are, on
average, about 0.6\,mag fainter in absolute magnitude, corresponding
to typical diameters roughly 25\,\% smaller. Nevertheless, the vast
majority of the new models still satisfy $H\leq12$, the limit adopted
by \cite{Han.ea:23} to minimize possible YORP-induced modification of
primordial spin states. Restricting the present sample to this
brightness range changes the prograde-to-retrograde ratio only
marginally, indicating that the weaker asymmetry cannot be explained
solely by the inclusion of smaller, potentially YORP-evolved bodies.
The prograde excess reported by \cite{Han.ea:23} therefore becomes
less pronounced when the sample is expanded from 90 to 247 unique
spin-state solutions.

Recently, \cite{Li.ea:25} derived rotation-period estimates for 216
Jupiter Trojans using photometric data from the ZTF survey. Of the 174
asteroids with reliability flags 3 or 2, we derived independent
rotation periods for 82 objects. Our periods agree with those of
\cite{Li.ea:25} in most cases but differ by more than 10\,\% for 26
objects. Several of the discrepant values correspond to approximate
factor-of-two aliases, while the largest differences preferentially
occur for slow rotators. This is plausibly related to the different
methodologies. The analysis of \cite{Li.ea:25} is based on
multiband Lomb--Scargle periodograms of ZTF photometry and inspection
of the resulting folded light curves, whereas our periods result from
a full convex inversion of all available photometry, often spanning
multiple apparitions and observing geometries. Sparse survey cadence
may make the recovery of very long periods difficult and can favor
shorter aliases, while a single-period analysis does not account for
the geometry-dependent light-curve changes included in the inversion
problem. The discrepant objects, therefore, merit a dedicated follow-up,
particularly those for which our models indicate periods of several
hundred hours.

\begin{figure}
        \includegraphics[width=\columnwidth]{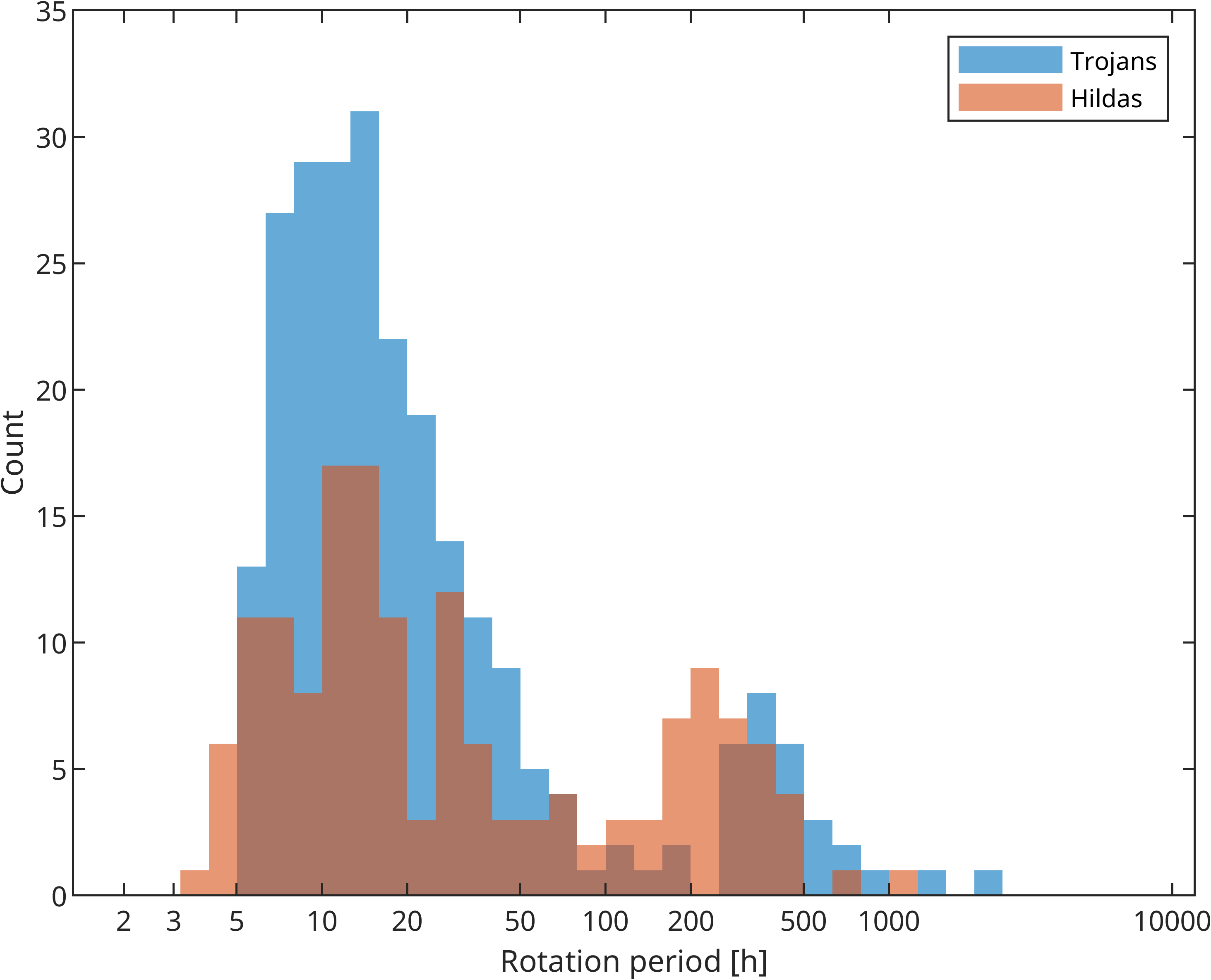}
        \caption{Period distribution of Jupiter Trojans and Hildas.}
        \label{fig:Trojans_Hildas_P}
    \end{figure}

We also recover the bimodal rotation-period distribution reported by
\cite{Kal.ea:21}, with a separation between the main population and
very slow rotators at approximately 100--200\,h
(Fig.~\ref{fig:Trojans_Hildas_P}). For main-belt asteroids, an excess
of slow rotators, particularly among small bodies, can be associated
with YORP-driven spin evolution. Such an explanation is less
plausible for the comparatively large and distant Jupiter Trojans
due to their much longer YORP timescales. An alternative proposed
by \cite{Nes.ea:20} is that the very slow rotators were originally
members of nearly equal-sized binary systems that tidally evolved
toward synchronous rotation. The companion could subsequently have
been either stripped during encounters with the giant planets prior to
Trojan capture or removed by later collisions, resulting in a slowly
rotating, single body. The observed excess of very slow Trojan
rotators may therefore preserve information about the primordial
binary fraction and the early dynamical evolution of their source
population. Slow rotators in our sample are typically more elongated,
although the difference from the remaining population is not
statistically significant.
    
    \subsection{Hildas}

\begin{figure}
        \includegraphics[width=\columnwidth]{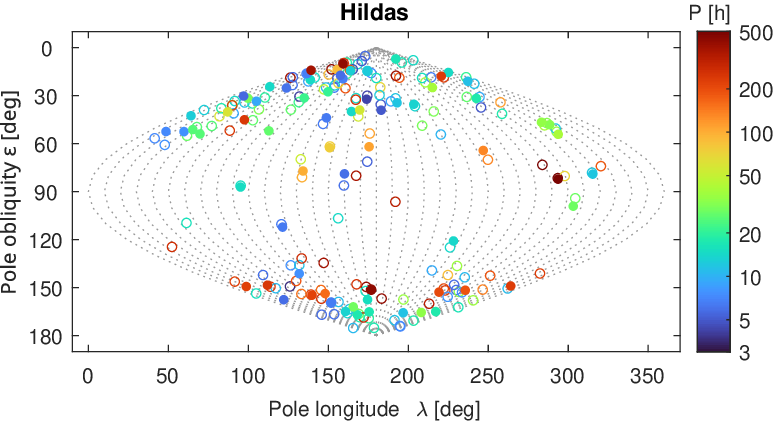}
            \caption{Spin and period distribution of Hildas. The filled circles represent full models (the first pole), and the empty circles are partial models (the best pole). The color of the circles corresponds to the rotation period. Rotation periods longer than 500\,h are mapped to 500\,h.}
        \label{fig:Hildas}
    \end{figure}

    The spin distribution of Hilda asteroids is shown in Fig.~\ref{fig:Hildas}. The clustering toward extreme values of obliquity is more prominent than for Trojans. The number of prograde Hildas in our sample was 94, and the number of retrograde Hildas 63. This is a statistically significant difference (as the null hypothesis -- that the true distribution is symmetric -- has a probability of 0.1\,\%). Nonetheless, the presence of possible selection effects may bias the distribution.

    The distribution of rotation periods (Fig.~\ref{fig:Trojans_Hildas_P}) is similar to that of Trojans, as previously shown by \cite{Sza.ea:20b}. The relative abundance of slow rotators with periods longer than 100\,h seems higher for Hildas than for Trojans.
    
    \subsection{Tumblers}

    The light curve inversion code we used to derive the models assumes that the asteroid's rotation axis is fixed and its rotation period is constant. However, some asteroids do not follow this assumption, as they rotate in an excited rotation state. Because our model cannot correctly describe their complex rotation, we expect that, in most such cases, the processing pipeline produces no model. However, there are cases in which the asteroid is known to be a tumbler, yet the pipeline produced a principal-axis-rotation model that passed all checks. 

\begin{figure}
        \includegraphics[width=\columnwidth]{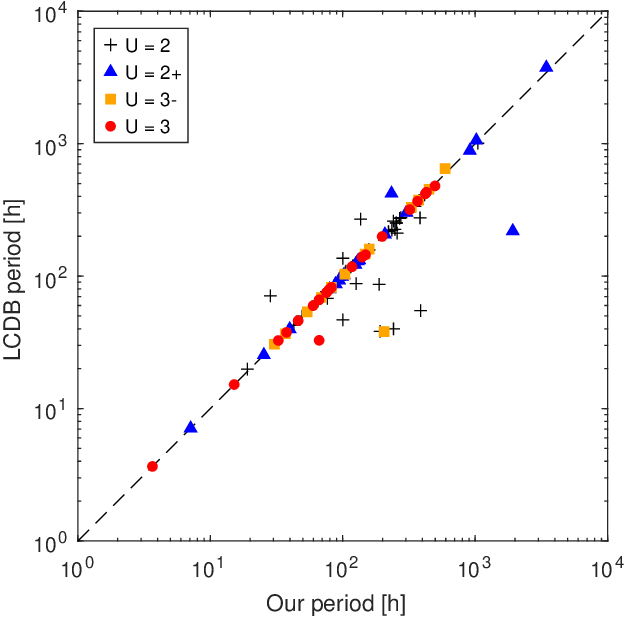}
        \caption{Comparison between the periods derived from our models and the periods of tumblers in the LCDB. The different symbols represent the different LCDB uncertainty codes. The diagonal dashed marks the exact match between the two periods.}
        \label{fig:tumblers}
    \end{figure}

    We selected those asteroids for which the LCDB marks them as tumblers, that is, those with flags T, T?, and T+. The comparison between our period and that in the LCDB is shown in Fig.~\ref{fig:tumblers} for 97 asteroids. For most tumblers, especially those with a reliable LCDB period with $U = 3$ or $U = 3-$, the periods are the same. This means that some of our models are tumblers. However, we modeled them as principal-axis rotators. If excitation is not high, the principal-axis model reproduces the main periodicity in the light curve and provides a satisfactory fit. We cite two examples. The first case consists of asteroid (5247)~Krylov -- with a main light-curve period of 82.28\,h and an angular-momentum spin vector at $(298^\circ, -58^\circ)$ \citep{Lee.ea:20} -- in which our model yields  $P = 82.2929$\,h  and a spin axis direction of $(278^\circ, -52^\circ)$. The second case consists of asteroid (319)~Leona, with a main period of 433\,h and an angular momentum vector of $(88^\circ, 11^\circ)$ \citep{Dur.ea:25}. We reconstructed two models with $P = 429.41$\,h. One pole solution was $(88^\circ, 18^\circ)$.

    This means that some of the asteroids that passed all our checks undergo tumbling, even though we successfully modeled them as principal-axis rotators. This is also crucial for interpreting the rotation periods in Fig.~\ref{fig:P_vs_D}, as the theoretical model of \cite{Zho.ea:25} predicts that most slow rotators are tumbling. However, tumbling can either be subtle -- with low excitation and minimal impact on light curves -- or substantial, with high excitation and complex rotation. Low excitation is difficult to detect even with densely sampled light curves \citep{Hen.Pra:13}. 
    
    \begin{figure*}
        \sidecaption
        \includegraphics[width=12cm]{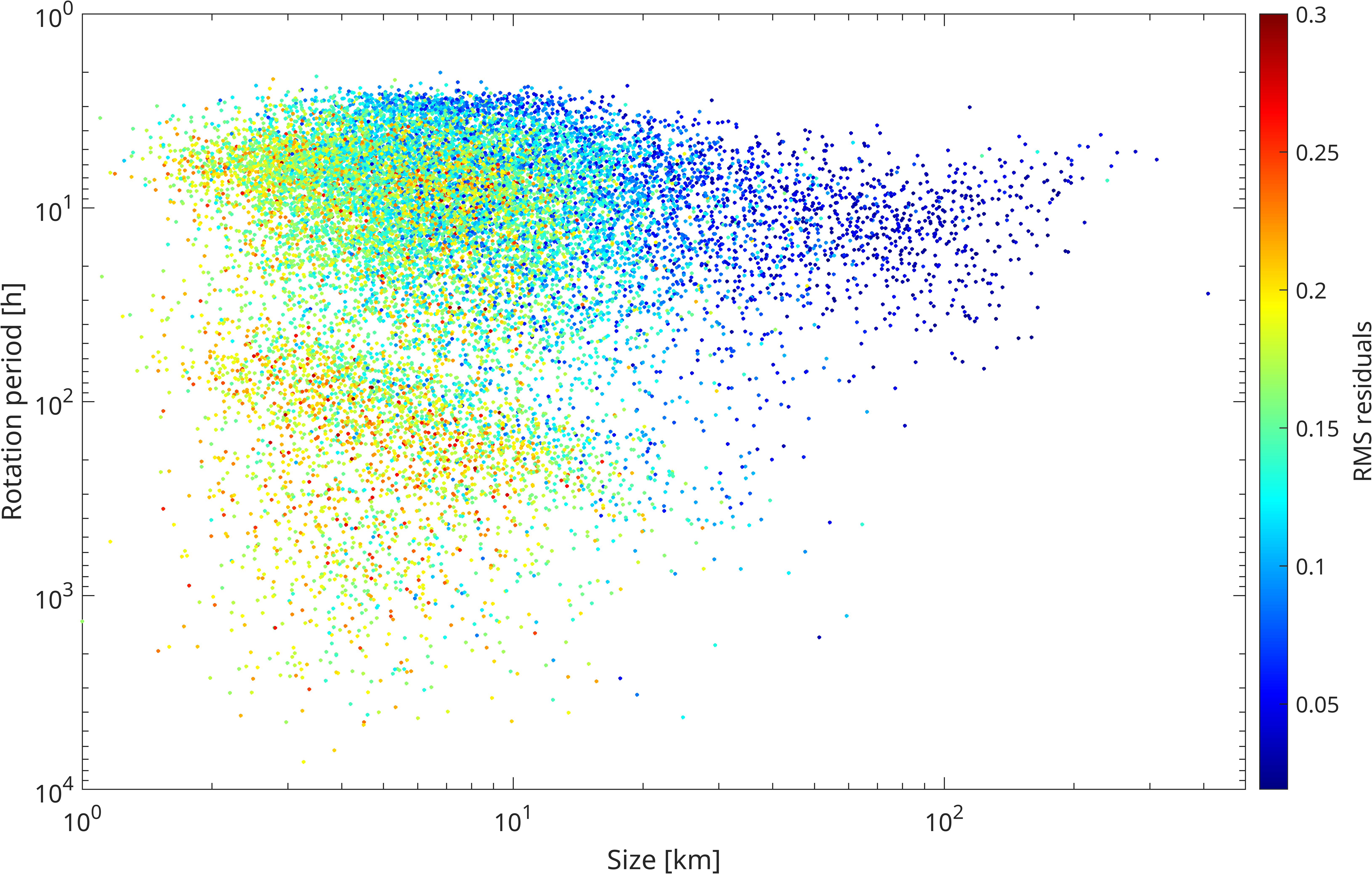}
        \caption{Period vs. size for convex shape models with available diameters. The color represents the RMS residual for ATLAS photometry in the $o$ filter. Slowly rotating asteroids below the valley have significantly larger residuals, likely because they are tumbling, so the model cannot fully fit their complex light curves.}
        \label{fig:P_vs_D_RMS}
    \end{figure*}

    Testing the prediction of \cite{Zho.ea:25} would require detailed modeling of a significant sample of slow rotators based on dense light curves. However, if slow rotators are mostly tumblers, we should see larger values of residuals than for the group of principal-axis rotators. Figure~\ref{fig:P_vs_D_RMS} shows the diameter--period plot (similar to Fig.~\ref{fig:P_vs_D}) for convex models. The color indicates the RMS residuals between the best-fit convex model and ATLAS $ o$-filter photometry. There is an expected horizontal trend in color: larger bodies are bluer, and smaller bodies are redder. This is because larger bodies are brighter, and their photometry is more accurate, leading to smaller residuals after the fit than for smaller, fainter bodies with larger photometric errors. There is also a vertical color trend: slow rotators below the valley contain more red points and fewer blue points than the faster rotators above. This is consistent with the prediction of \cite{Zho.ea:25} that slow rotators are mostly tumblers. Their model assumes principal-axis rotation, so it fits only the main period in the data. The residuals remain large because of tumbling and complex light curves, even for low-noise photometry.

    We chose the ATLAS $o$ filter for this test because it was available for the largest number of the asteroids in our sample (Table~\ref{tab:observatories}). However, the ATLAS $c$ filter, ZTF, or Gaia data qualitatively provide the same result.
    
    \subsection{Binaries}

    \begin{figure}
        \includegraphics[width=\columnwidth]{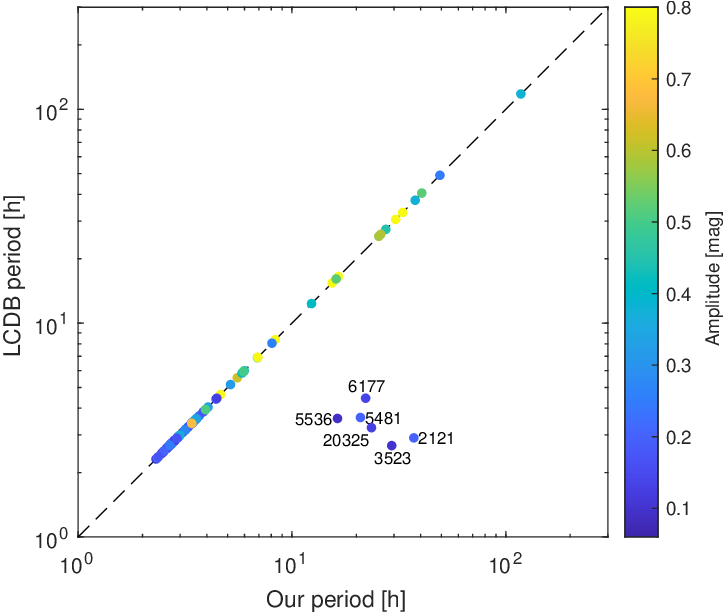}
        \caption{Comparison between the periods derived from our models and the LCDB primary periods of binary asteroids. The diagonal dashed line represents equality between the two periods. The color codes the amplitude of the LCDB period. Asteroids that do not lie on the diagonal are marked by their numbers. The period we detected corresponds to the orbital period of the secondary.}
        \label{fig:binaries_LCDB}
    \end{figure}

    Our sample also includes binary asteroids, which represent a significant fraction of the asteroid population. We selected all binaries (B tag) from the LCDB with U = 3 and amplitude information. There are 104 models shown in Fig.~\ref{fig:binaries_LCDB}. For most of these, the period obtained from our model agrees with the period reported in the LCDB for the primary. These are cases where either the signal is dominated by the primary and the secondary's effect is negligible, so it does not significantly affect the model, or the binary is fully synchronous and behaves as a single body from a modeling perspective. Six points that do not lie on the diagonal correspond to binaries for which the period obtained from our model equals the orbital period of the secondary. All these cases have a small amplitude for the primary period, so our algorithm picked the second period. The corresponding shape models do not correspond to the shapes of the primary or secondary; these are simply arbitrary shapes that fit the mutual events. 
    
\begin{figure}
        \includegraphics[width=\columnwidth]{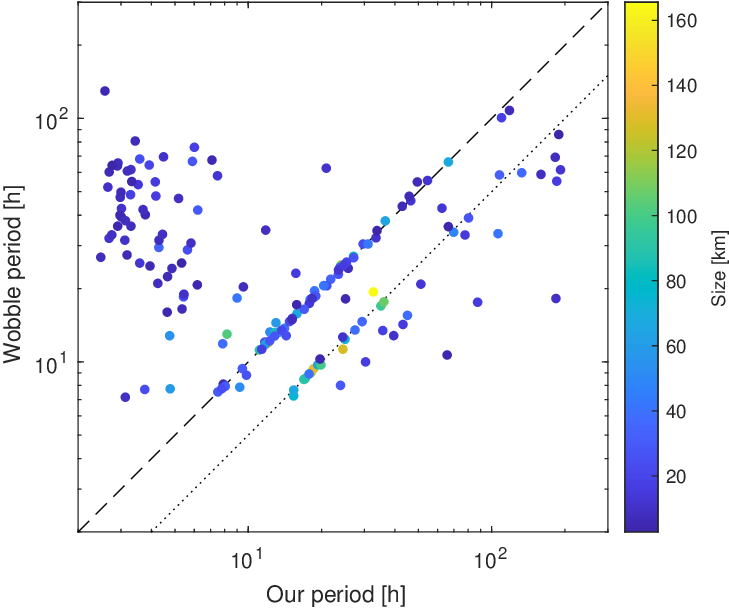}
        \caption{Comparison between the periods derived from our models and the wobble periods derived from Gaia binary candidates \citep{Lib.ea:26}. The dashed line represents equality between the two periods; the dotted line represents our rotation period, which is twice the wobble period. The color codes the asteroid size.}
        \label{fig:binaries_Gaia}
    \end{figure}

    We also compared our period determinations with binary candidates from Gaia astrometry by \cite{Lib.ea:26}, which is an update of \cite{Lib.ea:24}. From their list of 343 candidates, we selected 317 with $ p$ values below 5\,\%. For 170 of them, we obtained the period determination and provide a comparison in Fig.~\ref{fig:binaries_Gaia}. Many points lie on the main diagonal. These likely correspond to cases in which we detected the orbital period of the secondary or to double-synchronous binaries. As noted by \cite{Lib.ea:26}, the wobble detected for larger asteroids may in fact be a detection of the photocentric offset of a single body. The cluster of points above the diagonal may indicate the rotation period of the primary, while the wobble period represents the orbital period of the secondary.

  \section{Conclusions}

  By deriving new convex-shape models of asteroids, we increased their number from 10,000 to about 25,000. Together with almost 10,000 partial models, this represents a large dataset for studying asteroid spin properties across populations.  The current sample shows a complex picture of the distribution of rotation periods and their correlation with shape: the fastest rotators near the spin barrier are mostly spheroidal, because the spin barrier shifts toward longer periods for more elongated objects \citep{Pra.Har:00}. Another factor affecting the value of the critical spin period is density. The S- and C-type asteroids have a different distribution of rotation periods, not only near the spin barrier but also at longer rotation periods. The position of the valley separating fast and slow rotators differs for these two taxonomic classes \citep{Dye.ea:26}. Most of the slow rotators below the valley might be, in fact, tumblers, as predicted by \cite{Zho.ea:25} and supported by the fact that photometry of these asteroids exhibits larger residuals when fitted with a model assuming principal-axis rotation.

  \section{Data availability}
  
  All new models are available through the DAMIT and at Zenodo \url{https://doi.org/10.5281/zenodo.22812438}. In addition to the shapes and spin parameters, we also provide color parameters, defined as offsets between filters, and phase-function parameters.

    \begin{acknowledgement}
        This work has been supported by the grant 23-04946S of the Czech Science Foundation. The results presented here would not have been possible without the hundreds of thousands of volunteers who freely provided computing resources for the Asteroids@home distributed computing project. We greatly appreciate their contribution and patience.
        This research has made use of IMCCE's Miriade VO tool and the JPL Horizons online ephemeris computation service.
        This work uses data from the University of Hawaii's ATLAS project, funded through NASA grants NN12AR55G, 80NSSC18K0284, and 80NSSC18K1575, with contributions from the Queen's University Belfast, STScI, the South African Astronomical Observatory, and the Millennium Institute of Astrophysics, Chile.
        This work has made use of data from the European Space Agency (ESA) mission {\it Gaia} (\url{https://www.cosmos.esa.int/gaia}), processed by the {\it Gaia} Data Processing and Analysis Consortium (DPAC, \url{https://www.cosmos.esa.int/web/gaia/dpac/consortium}). Funding for the DPAC has been provided by national institutions, in particular the institutions participating in the {\it Gaia} Multilateral Agreement.
        This work is partly based on data provided by the Minor Planet Physical Properties Catalogue (MP3C) of the Observatoire de la Côte d'Azur. 
    \end{acknowledgement}

    \bibliographystyle{aa}
    \bibliography{pepa}

\onecolumn

\begin{appendix}

  \section{Different DAMIT periods}
    \begin{longtable}{rddcc|rddcc}
      \caption{\label{tab:DAMIT} Different DAMIT periods.}\\
        \hline \hline
        Asteroid    & \multicolumn{1}{c}{$P$ [h]}   & \multicolumn{1}{c}{$P_\mathrm{DAMIT}$ [h]}      & Note   & Ref. & Asteroid    & \multicolumn{1}{c}{$P$ [h]}   & \multicolumn{1}{c}{$P_\mathrm{DAMIT}$ [h]}      & Note   & Ref.\\
        \hline
      \endfirsthead
      \caption{continued.}\\
      \hline\hline
      Asteroid    & \multicolumn{1}{c}{$P$}   & \multicolumn{1}{c}{$P_{DAMIT}$}      & Note   & Ref. & Asteroid    & \multicolumn{1}{c}{$P$}   & \multicolumn{1}{c}{$P_{DAMIT}$}      & Note   & Ref.\\
      \hline
      \endhead
      \hline
      \endfoot
10 & 27.6589 & 13.8256 & D, R  & (9)  &    10064 & 8.05352 & 12.1277 & & (6)    \\
213 & 24.0726 & 12.0395 & D & (4)   &             10579 & 16.4747 & 33.1577 & & (2)    \\
408 & 405.239 & 202.62 & D & (8)   &              10672 & 4.90459 & 3.26966 & & (4)    \\
410 & 32.4954 & 32.9961 &  & (2)   &                   11168 & 1457.66 & 49.447 &   & (3)   \\
449 & 36.5484 & 18.0386 & D & (3)   &             11184 & 4.15021 & 4.06117 &  & (4)   \\
954 & 7.20519 & 14.4098 & H & (4)   &               11235 & 20.2213 & 20.5136 &  & (2)   \\
988 & 621.045 & 947.7 & & (8)    &                     11329 & 4.4648 & 4.03319 & & (4)    \\
1170 & 3.13611 & 15.9295 & & (1)    &                  11497 & 18.8686 & 9.435 & D & (4)   \\
1465 & 4.88138 & 11.949 & & (2)    &                   11618 & 33.7023 & 40.2337 & & (1)    \\
1510 & 5.86712 & 15.9844 & & (1)    &                  12226 & 7.15402 & 7.5468 & & (4)    \\
1665 & 162.63 & 67.9106 & & (6)   &                    12396 & 10.4992 & 4.45005 & & (2)    \\
1702 & 21.141 & 22.801 &  & (4)   &                    12729 & 10.8008 & 14.6011 & & (2)    \\
1957 & 3.67346 & 3.79361 & & (2)    &                  12809 & 57.3783 & 45.336 & & (4)    \\
2180 & 7.7031 & 8.34623 & & (7)    &                   12883 & 5.84516 & 12.4933 & & (2)    \\
2234 & 23.6055 & 15.8124 & & (8)    &                  12891 & 14.59 & 8.04341 & & (2)    \\
2364 & 4.41219 & 24.2004 & & (8)    &                  13060 & 306.869 & 482.46 & & (2)    \\
2367 & 576.584 & 564.4 &   & (4)   &                   13116 & 47.2602 & 37.1773 & & (2)    \\
2393 & 7.91574 & 9.28753 & & (6)    &                  13289 & 438.202 & 43.2618 & & (2)    \\
2791 & 16.3495 & 9.80729 & & (6)    &                  13585 & 100.563 & 50.28 & D & (1)   \\
2962 & 2.67804 & 2.53632 & & (6)    &                  13612 & 5.17267 & 2.25519 & & (4)    \\
3093 & 1119.32 & 432.34 & & (2)    &                   13617 & 14.7878 & 20.0635 & & (2)    \\
3097 & 7.41701 & 8.77591 & & (5)    &                  14555 & 35.6527 & 23.856 & & (2)    \\
3140 & 32.4941 & 43.2235 & & (2)    &                  15276 & 12.9685 & 13.3299 & & (1)    \\
3233 & 865.275 & 886.7 &   & (8)   &                   15735 & 21.8538 & 22.5355 & & (1)    \\
3476 & 7.0191 & 2.5131 & & (4)    &                    15847 & 1883.73 & 46.798 &   & (3)   \\
3611 & 31.908 & 31.2783 &   & (1)   &                  16009 & 16.691 & 8.34757 & D & (5)   \\
3847 & 365.475 & 360 &   & (2)   &                     16025 & 10.8526 & 10.5573 &   & (1)   \\
4400 & 2.54695 & 9.3275 & & (4)    &                   16394 & 3.43226 & 26.8984 & & (2)    \\
4437 & 77.2986 & 29.5755 & & (8)    &                  16712 & 17.6566 & 17.1699 &   & (1)   \\
4488 & 78.7359 & 77.937 &   & (2)   &                  16785 & 282.611 & 16.9409 & & (2)    \\
4714 & 14.7505 & 12.4253 & & (2)    &                  17162 & 6.16061 & 2.93962 & & (4)    \\
4770 & 35.1025 & 68.28 & & (4)    &                    18414 & 52.5989 & 25.073 & & (3)    \\
4846 & 16.3789 & 20.035 & & (2)    &                   18591 & 2.67003 & 7.03654 & & (2)    \\
5008 & 27.1649 & 49.238 & & (6)    &                   18627 & 4.46419 & 39.507 & & (4)    \\
5372 & 1297.7 & 68.218 &   & (2)   &                   18897 & 12.5613 & 12.7378 &   & (2)   \\
5497 & 1373.25 & 46.284 &   & (3)   &                  19274 & 127.074 & 165.95 & & (4)    \\
5589 & 4.17223 & 13.517 & & (2)    &                   21181 & 115.118 & 72.957 & & (2)    \\
5746 & 270.188 & 267.21 &   & (8)   &                  22337 & 10.0533 & 9.4974 & & (4)    \\
6026 & 4.10587 & 3.7817 & & (6)    &                   22473 & 19.6026 & 3.97838 & & (4)    \\
6051 & 2.93881 & 38.911 & & (1)    &                   23142 & 2.39798 & 21.6658 & & (1)    \\
6104 & 26.7498 & 30.89 & & (4)    &                    23836 & 5.61487 & 2.80758 & D & (4)   \\
6286 & 26.4004 & 28.0609 & & (2)    &                  24210 & 10.2844 & 2.64752 & & (4)    \\
6349 & 4.37613 & 23.8123 & & (8)    &                  24971 & 4.3072 & 39.8999 & & (1)    \\
6683 & 10.8367 & 14.0056 & & (3)    &                  25820 & 11.9356 & 12.0589 &   & (1)   \\
7479 & 4.92499 & 2.03659 & & (4)    &                  26473 & 11.2003 & 10.974 &   & (4)   \\
7632 & 6.44056 & 5.29072 & & (5)    &                  26593 & 3.83306 & 3.66189 & & (4)    \\
7779 & 4.38014 & 2.19003 & D & (4)   &            26706 & 355.688 & 420.6 & & (4)    \\
7815 & 57.6948 & 3.66589 & & (4)    &                  26854 & 3.12305 & 13.4742 & & (2)    \\
7859 & 10.166 & 10.7035 &   & (1)   &                  26966 & 4.96148 & 3.45463 & & (4)    \\
8043 & 946.103 & 22.7606 & & (6)    &                  27070 & 5.47729 & 4.91443 & & (3)    \\
8866 & 13.7076 & 3.91153 & & (4)    &                  28146 & 14.364 & 7.1819 & D & (4)   \\
9112 & 34.4507 & 36.537 & & (1)    &                   28362 & 36.53 & 39.4903 & & (2)    \\
9119 & 11.695 & 2.94707 & & (4)    &                   28630 & 46.2062 & 4.25443 & & (4)    \\
9165 & 1303.39 & 46.22 &   & (3)   &                   29684 & 226.298 & 436 & H & (4)   \\
9171 & 14.0263 & 13.6672 & & (2)    &                  30224 & 16.2054 & 47.8972 & & (2)    \\
9480 & 220.687 & 218.2 &   & (4)   &                   31193 & 70.9503 & 73.05 &   & (4)   \\
9542 & 3.36478 & 2.79471 & & (6)    &                  31753 & 14.8248 & 14.6636 &   & (1)   \\
9712 & 32.7269 & 16.3651 & D & (4)   &            32310 & 2.99463 & 2.93302 &   & (4)   \\
33714 & 413.56 & 218.1 & D & (4)   &              64192 & 6.68555 & 7.3166 & & (4)    \\  
35673 & 34.7559 & 17.376 & D & (4)   &                 67005 & 24.0068 & 8.32667 & & (1)    \\
36572 & 6.2289 & 7.01203 & & (2)    &                  75173 & 5.89841 & 2.94918 & D & (4)   \\
36765 & 16.1846 & 13.2817 & & (4)    &                 81878 & 4.60212 & 2.30107 & D & (4)   \\
37586 & 911.992 & 50.558 & & (3)    &                  85128 & 3.30702 & 62.23 & & (2)    \\
40429 & 399.331 & 403.9 &   & (8)   &                  86401 & 78.0752 & 29.721 & & (3)    \\
41161 & 5.67122 & 2.60366 & & (4)    &                 88031 & 38.3849 & 4.46337 & & (4)    \\
42923 & 7.14142 & 8.38888 & & (6)    &                 92064 & 6.27997 & 11.0048 & & (4)    \\
43987 & 12.2675 & 12.0188 &   & (1)   &                96155 & 15.5818 & 17.6041 & & (1)    \\
44878 & 91.856 & 53.236 & & (4)    &                   96513 & 13.3793 & 6.68802 & D & (4)   \\
45386 & 142.137 & 328.9 & & (4)   &                    109214 & 5.75272 & 6.81069 & & (1)    \\
45837 & 62.7601 & 83.56 & & (4)    &                   111029 & 198.392 & 63.104 & & (1)    \\
45883 & 8.70004 & 8.0052 & & (4)    &                  114773 & 10.5746 & 10.4421 &   & (4)   \\
46244 & 5.0138 & 4.87787 & & (2)    &                  115216 & 5.30276 & 2.54669 & & (4)    \\
48007 & 169.902 & 134.573 & & (2)    &                 115755 & 7.78943 & 7.59219 &   & (1)   \\
50776 & 11.1478 & 16.7993 & & (1)    &                 118762 & 224.486 & 3.66931 & & (4)    \\
51832 & 19.7585 & 20.8929 & & (1)    &                 131466 & 8.01256 & 8.197 &   & (4)   \\
52809 & 6.45367 & 2.4916 & & (4)    &                  138831 & 7.7728 & 2.28689 & & (1)    \\
52998 & 6.03983 & 5.81972 & & (4)    &                 143173 & 13.9175 & 3.30715 & & (1)    \\
54355 & 6.94548 & 2.30616 &  & (4)   &                 162515 & 264.1 & 131.97 & D & (4)   \\
56256 & 82.6728 & 56.324 & & (4)    &                  163430 & 48.2515 & 13.6396 & & (1)    \\
56433 & 9.88316 & 40.994 & & (4)    &             164344 & 174.463 & 151.09 & & (4)    \\
59336 & 7.16691 & 3.58326 & D & (4)   &           189260 & 6.84111 & 5.99141 & & (1)    \\
61253 & 96.0388 & 50.87 & & (4)    &            & & & & \\

    \end{longtable}
     \tablefoot{The table lists all asteroids for which the relative difference between our period $P$ and the period in DAMIT $P_\mathrm{DAMIT}$ was larger than 1\%. In some cases, our new period was twice the DAMIT period, which we denote as D -- double. In two cases, our period was half (H) the DAMIT period. The model of asteroid (10)~Hygiea was rejected (R) because its light curve is dominated by albedo effects, which our method does not take into account.}
\noindent \small {\bf References.} (1) \cite{Dur.ea:18c}; (2) \cite{Dur.ea:19}; (3) \cite{Dur.ea:20}; (4) \cite{Dur.Han:23}; (5) \cite{Han.ea:13b}; (6) \cite{Han.ea:16}; (7) \cite{han.ea:18a}; (8) \cite{Han.ea:21}; (9) \cite{Ver.ea:20}.      

\end{appendix}

\end{document}